\documentclass[twocolumn]{aastex6}


\usepackage{amsmath}
\usepackage{graphicx}
\usepackage{dcolumn}
\usepackage{bm}
\usepackage{comment}
\begin{document}



\title{Redshift Evolution of the Black Hole Merger Rate from Globular Clusters}

\author{Carl L.\ Rodriguez\altaffilmark{1} and Abraham Loeb\altaffilmark{2}}

\altaffiltext{1}{Pappalardo Fellow; MIT-Kavli Institute for Astrophysics and Space Research, 77 Massachusetts Avenue, 37-664H, Cambridge, MA 02139, USA}
\altaffiltext{2}{Astronomy Department, Harvard University, 60 Garden Street, Cambridge, MA 02138}

\date{\today}

\begin{abstract}
	As the sensitivity of current and future gravitational-wave detectors 
	improves, it will become possible to measure the evolution of the binary 
	black hole merger rate with redshift.  Here, we combine detailed 
	fits to state-of-the-art dynamical models of binary black hole formation 
	in dense star clusters with a cosmological model of cluster formation 
	across cosmic time.  We find a typical merger rate of 14 
	$\rm{Gpc}^{-3} \rm{yr}^{-1}$ in the local universe, with a reasonable 
	range of 4-18 $\rm{Gpc}^{-3} \rm{yr}^{-1}$, depending on the rate of 
	cluster disruption and the cluster initial mass function.   This rate 
	increases by a factor of 6 to redshift $z=2.7$ before declining at higher redshifts.  We compare the merger rate 
	from binaries produced in clusters to similar estimates from isolated binaries and triples in 
	galactic fields, and discuss various ways that these different formation 
	channels could add up to the current merger rate observed by LIGO/Virgo.
\end{abstract}

\maketitle


\section{Introduction}
\label{sec:level1}

The detection of gravitational waves (GWs) from merging binary black holes 
(BBHs) by Advanced LIGO/Virgo 
\citep{Abbott2017c,Abbott2017,Abbott2017d,Abbott2016a,Abbott2016} has stimulated 
many theoretical questions about their origin.  While a rich variety of formation 
channels have been proposed to explain these events, the vast majority fall into 
one of two categories.  In the first, the BBHs are formed as the end-stage of evolution for a massive binary, and merge through the emission of GWs either following a 
common-envelope phase \citep[the ``field'' channel, e.g.,][]{Belczynski2002,Voss2003,
Podsiadlowski2003,Sadowski2007a,Belczynski2010,Dominik2012,Dominik2014,Dominik2013,Belczynski2016},
or by secular interaction with a third companion\citep[the ``triple'' channel, e.g.,][]{Antonini2012a,Antonini2016,
VanLandingham2016,Leigh2017,Silsbee2017,Petrovich2017,Hoang2018,Antonini2017,Rodriguez2018ALIGO/Virgo}.
In the second category, BBHs are dynamically-forged through two- or three-body 
encounters in dense stellar environments such as globular clusters (GCs) or 
galactic nuceli \citep[e.g.,][]{PortegiesZwart2000,Banerjee2010,Ziosi2014,Banerjee2017,
OLeary2006,OLeary2007,Moody2009,Downing2010,Downing2011,Tanikawa2013,Bae2014,
Rodriguez2015a,Rodriguez2016a,Rodriguez2016b,Askar2016,Giesler2017,Rodriguez2018}.  

While any of these formation channels can produce BBH mergers with masses and 
spins similar to those observed by LIGO/Virgo, the physical processes that drive BBHs to merge operate on significantly different timescales in each channel.  Even though the majority of these mechanisms are modulated by 
the same cosmic star-formation rate (SFR), the different delay times between BBH 
formation and merger will produce different merger rate distributions in each over redshift.  These differences may be detectable by either the 
current \citep[][]{Fishbach2018} or future \citep[][]{Abbott2017f,Hild2011,Vitale2018} generation of GW detectors, and can be used 
to disentangle the contributions of different formation channels to the overall 
BBH merger rate.

In this letter, we use state-of-the-art dynamical models of GCs from 
\cite{Rodriguez2018} and a detailed model of GC formation across cosmic time 
\citep{El-Badry2018} to compute a cosmological rate of BBH mergers from the 
dynamical channel.  Unlike previous calculations \citep[][]{PortegiesZwart2000,Rodriguez2016a,Askar2016}, this calculation allows us to 
directly compare the evolution of the dynamical BBH merger rate to that from other 
channels shaped by the cosmic SFR, such as the classical field channel 
\citep[taken from][]{Belczynski2016} and the field triple channel 
\citep[taken from][]{Rodriguez2018ALIGO/Virgo}.

In Section \ref{sec:ratefits}, we describe the details of our dynamical GC models, and how we use the GC formation model of \cite{El-Badry2018} to compute the BBH merger rate.  In Section \ref{sec:3}, we explore the evolution of the merger rates over redshift, and show how our model depends on assumptions about GC disruption and the cluster initial mass function (CIMF).  Finally, in Section \ref{sec:rates}, we compare the cosmic merger rates from GCs to those from isolated binaries and from stellar triples, and show how the current LIGO/Virgo merger rates can be explained by different combinations of the three different formation channels.  Throughout this paper, we assume a $\Lambda$CDM cosmology with $h= 0.679$ and $\Omega_M = 0.3065$ \citep{PlanckCollaboration2015}. 

  \section{Rate Fits}
  \label{sec:ratefits}

\cite{El-Badry2018} have developed a formalism to model the formation 
of GCs by populating galaxy halos with GCs based on gas mass and pressure, while tracking the fate of these GCs during the merger assembly history of their host galaxies.  To get the formation of GCs in different halo masses 
across cosmic time, we created phenomenological fits to their total SFR in GCs per comoving volume per halo mass at a 
given redshift.  We followed their ``standard'' model, but included the 
additional factor of 2.6 in their $\alpha_\Gamma$ parameter to account for 
cluster disruption (see Appendix \ref{app:elbadry}).  To translate this 
into a rate of BBH mergers, we need only to convolve this cosmological cluster SFR 
with the rate of BBH mergers from GCs.  Our rate equation for the merger rate of BBHs at a cosmic time $t$ is:

  \begin{align}
 \mathcal{R}(t) = &\iiint \left. \frac{ \dot{M}_{\rm GC}} {d\log_{10}M_{\rm    
	 Halo}} \right|_{z(\tau)} \frac{1}{\left< M_{\rm GC}\right>} 
	 P(M_{\rm GC}) \nonumber\\
	 & \times R(r_{\rm v},M_{\rm GC},\tau-t) dM_{\rm{Halo}}dM_{\rm GC} d\tau~,
 \label{eqn:master}
  \end{align}
  
  \noindent where  $\frac{ \dot{M}_{\rm GC}} {d\log_{10}M_{\rm    Halo}}$ is the 
	      comoving rate of star formation in GCs (in units of $M_{\odot} {\rm yr}^{-1} {\rm Mpc}^{-3} $) per 
	      galaxies of a given halo mass $M_{\rm Halo}$ at a redshift $z(\tau)$ corresponding to a formation time $\tau$.  
       $P(M_{\rm GC})$ is the cluster initial mass function (CIMF), which 
	      we take to be $\propto M_{\rm GC}^{-2}$ between $10^5M_{\odot}$ 
	      and $10^7 M_{\odot}$ (though we explore variations to this in 
	      Section \ref{sec:3}).
       $\left< M_{\rm GC}\right>$ is the mean initial mass of a GC given 
	      our assumed CIMF.  This converts the mass that goes into forming GCs into 
	      the number of GCs formed.
       Finally, $R(r_{v},M_{\rm GC},t)$ is the rate of mergers (ejected and 
	      in-cluster) for a GC with a given initial virial radius $r_v$ and 
	      mass $M_{\rm GC}$ at a time $t$ after its formation.

  For our merger rate from individual clusters, there are two different 
  effects which must be considered.  First, as noted in \cite{Rodriguez2016a}, 
  the dependence of the BBH merger rate on cluster mass is super-linear.  
  While the number of BHs in any given cluster scales linearly with the cluster 
  mass, the fraction of BBHs that will merge in a Hubble time 
  also increases with the mass of the cluster.  This occurs because a more 
  massive cluster forms BBHs in a deeper potential well, requiring more 
  scattering encounters to harden the binary until its inevitable ejection or merger.  At the same time, as the cluster loses mass and expands over time, the rates of 
  BBH hardening and ejection will decrease.  This causes the BBH merger rate to 
  decrease exponentially over time  \citep[e.g.,][]{Tanikawa2013,Hong2018}.  Combining these 
  different physical intuitions, we find that the following phenomenological 
  rate formula provides a good fit to the BBH merger rate of individual 
  cluster:  
  \begin{equation}
	  R(M_{\rm GC},t) \equiv  (A M_{\rm GC}^2 + B M_{\rm GC} + 
	  C) \times t^{-(\gamma + \gamma_M \log_{10}M_{\rm GC})}
	  \label{eqn:gcrate}
  \end{equation}

  \noindent where $M_{\rm GC}$ is the initial cluster mass and $t$ is the age of 
  the given cluster.  Since the models from \cite{Rodriguez2018} 
  only cover two virial radii 
  ($1\rm{pc}$ and $2\rm{pc}$), we do not attempt to incorporate this information in our 
  fit \citep[though see][for an exploration of the parameter space of cluster initial radii]{Hong2018,Choksi2018a}.  Instead, we separately fit \eqref{eqn:gcrate} to all the models with 
  $r_v = 1\rm{pc}$ and $r_v = 2\rm{pc}$.  In reality, our model should 
  incorporate information about the virial radius in the fit itself \citep[as was done using the central 
  cluster density in ][]{Hong2018}, but the separate fits allow us to 
  disentangle the influence of cluster concentration on the redshift 
  distribution of BBH mergers.  We also fit separately the 
  in-cluster mergers and those that are ejected from the cluster and merge later, since there is no reason to expect 
  them to follow the same phenomenological fits.  
  This produces 4 total sets of parameters $\theta = (A,B,C,\gamma,\gamma_M)$.  
  For more details about our fitting procedure, see Appendix \ref{app:rate}.

  \section{Cosmological Merger Rates}
  \label{sec:3}

\begin{figure}[t!]
\centering
\includegraphics[scale=0.85, trim=0.in 0.3in 0in 0.65in, clip=true]{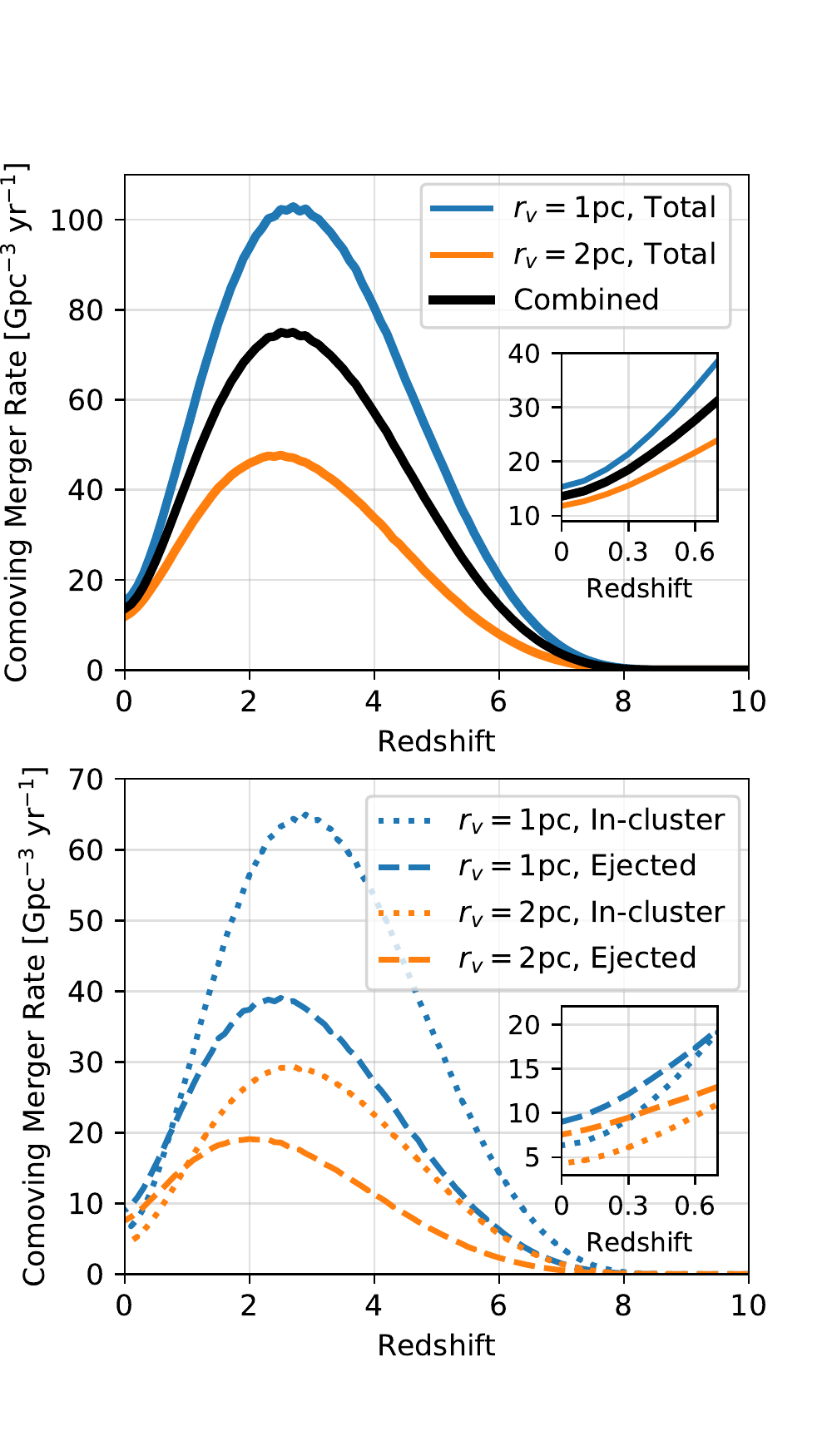}
\caption{Comoving Merger Rates of BBHs from GCs in our standard model.  We show separately the evolution of the in-cluster and ejected BBH mergers and models with initial virial radii of 1 and 2 pc. {Our fiducial model assumes 50\% of clusters form with $r_v = 1\rm{pc}$ and 50\% form with $r_v = 2\rm{pc}$.}}
\label{fig:ejin}
\end{figure}

  Figure \ref{fig:ejin} shows the standard merger rates as a function of 
  redshift using our phenomenological fits and equation \eqref{eqn:master}.  We 
  show separately the in-cluster and ejected 
  mergers for BBHs from clusters with $r_v = 1\rm{pc}$ and $r_v=2\rm{pc}$.  In all four 
  cases, the merger rate slowly increases as GCs are formed in the early 
  universe.  
  The in-cluster mergers peak at $z\sim 3$ ($z\sim 2.6$) for the $1\rm{pc}$ ($2\rm{pc}$) clusters, while the ejected 
  mergers peak later at $z\sim 2.3$ ($z\sim2$).  
  This delay is expected: in-cluster mergers are 
  prompt mergers, occurring almost immediately after the last dynamical encounter in the cluster.  
  Ejected mergers, on the other hand, can sometimes 
  experience a significant delay between their ejection and subsequent merger 
  \citep[as long as 10 Gyr, see][]{Rodriguez2016b}.  At the same time, 
  clusters with larger virial radii have correspondingly longer half-mass 
  relaxation times ($\propto r_v^{3/2}$).  Since the 
  time for the BHs to segregate near the cluster center also scales as 
  the half-mass relaxation time of the cluster, GCs with larger virial radii 
  will require more time to dynamically form BBHs; this, in turn, produces a 
  lag in both the ejected and in-cluster mergers.

  At early times, the in-cluster mergers dominate the BBH merger rates, with the 
  1pc models predicting a maximum in-cluster merger rate of $\gtrsim 
  60~\rm{Gpc}^{-3}\rm{yr}^{-1}$ at $z \sim 3$.  However, the delay time between 
  formation and merger for the ejected BBHs shifts the distribution towards lower redshifts, such that at late times the rate of ejected 
  BBHs is nearly twice that of the in-cluster mergers.  {For the 1pc mergers, the ejected mergers become dominant at $z\lesssim 0.7$, which increases to $z\lesssim 0.9$ for the 2pc mergers.}  The total merger rate 
  from our models in the local universe ($z \lesssim 0.1$) is $15~
  \rm{Gpc}^{-3}\rm{yr}^{-1}$ for GCs with $r_v = 1\rm{pc}$ and $12~\rm{Gpc}^{-3}\rm{yr}^{-1}$
  for $r_v = 2\rm{pc}$.  
  
  {As a general trend, a larger virial radius decreases the overall merger rate, but increases the delay between BBH formation and merger, flattening the distribution in the local universe.  This trend has been independently shown by \cite{Choksi2018a}, which used semi-analytic models of BBH mergers from GCs to fully explore the parameter space of initial conditions.  To limit our computational requirements, we have assumed virial radii of 1 and 2pc because this brackets the observed peak of effective radii for star formation in the local universe \citep[e.g.][]{Scheepmaker2007}.  Furthermore, recent theoretical work \citep{Kremer2018} has shown that very compact initial radii are required to eject the majority of BHs in any given GC, necessary to reproduce the observed surface brightness profiles of core-collapsed clusters.  For the remainder of this paper, we will assume a standard model where 50\% of clusters form with $r_v = 1\rm{pc}$ and 50\% form with $r_v = 2\rm{pc}$.  This yields a local merger rate of $14~
  \rm{Gpc}^{-3}\rm{yr}^{-1}$, where the ejected mergers become dominant for $z\lesssim 0.8$.}

  \subsection{Variations of the CIMF}

  As in \cite{El-Badry2018}, we have assumed that the GC initial mass 
  function follows a simple $M_{\rm GC}^{-2}$ distribution from 
  $10^5M_{\odot}$ to $10^7M_{\odot}$.  However, there is 
  observational evidence that the CIMF contains an exponential truncation 
  at higher masses, and that the truncation depends on galaxy type 
  \citep[see][for a review]{PortegiesZwart2010}.   This function 
  is often written as 

  \begin{equation}
	  \phi(M_{\rm GC})dM_{\rm GC} \propto M_{\rm GC}^{-2} \exp( - M_{\rm GC} / 
	  M_{\rm GC}^{\bigstar})dM_{\rm GC}~,
	  \label{eqn:schechter}
  \end{equation}

  \noindent where $M_{\rm GC}^{\bigstar}$ is the truncation mass.

\begin{figure}[t!]
\centering
\includegraphics[scale=0.85, trim=0.in 0.1in 0in 0.0in, clip=true]{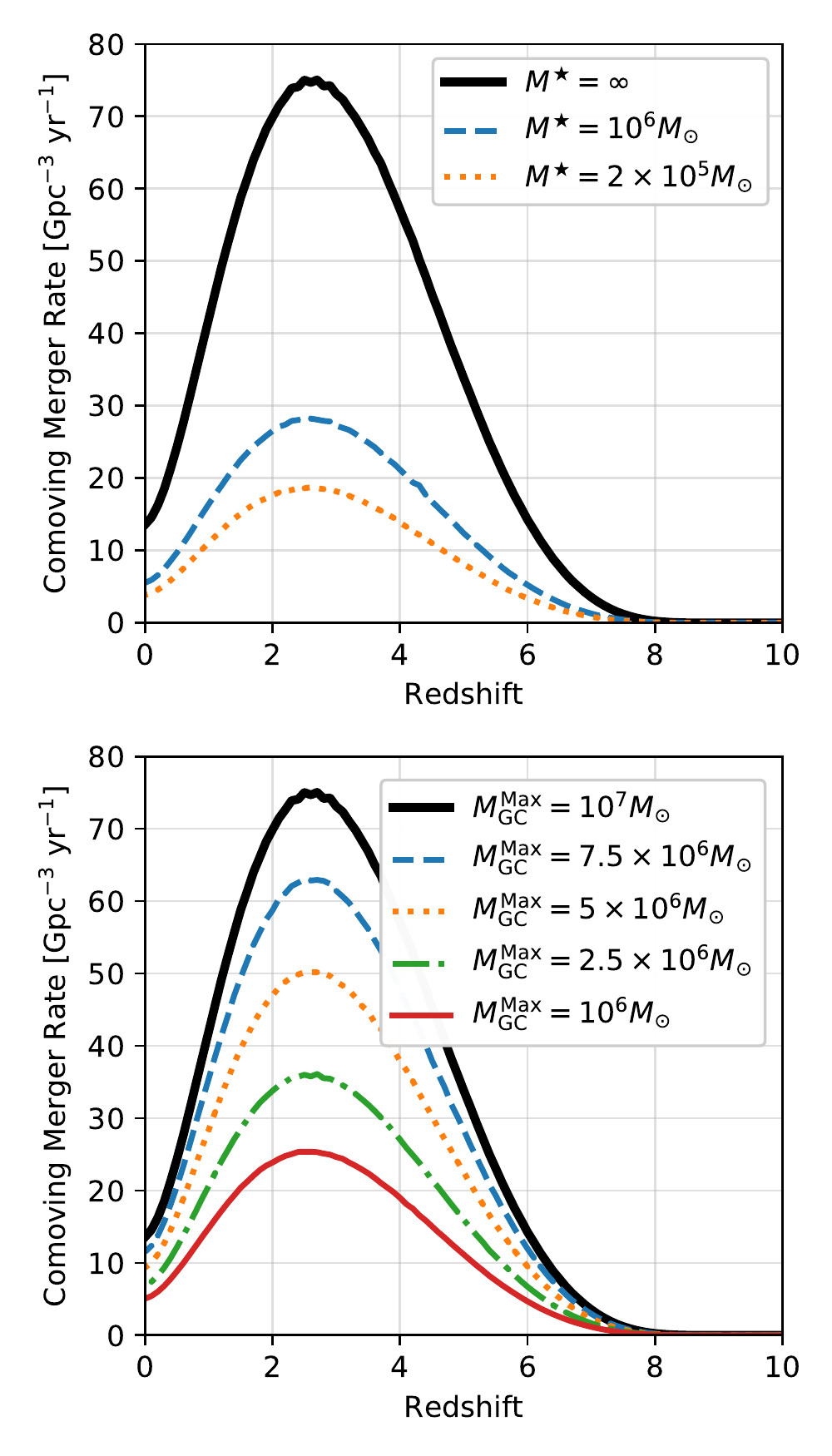}
\caption{The sensitivity of our result from Equation \eqref{eqn:master} to the cutoff of the CIMF.  In the top panel, we show how the introduction of an exponential truncation in the $M_{\rm GC}^{-2}$ mass function significantly reduces the merger rate.  In the bottom panel, we show how the rate reduces as a function of the maximum GC mass at formation.}
\label{fig:mstar}
\end{figure}

  To test the influence of the CIMF on our present estimate, we 
  recompute Equation \eqref{eqn:master} with $M_{\rm GC}^{\bigstar} = 2\times 
  10^5M_{\odot}$ \citep[as suggested by observations for spiral galaxies,][]{Gieles2006,Larsen2009} and 
  $M_{\rm GC}^{\bigstar} = 10^6 M_{\odot}$ \citep[as suggested by observations for 
  interacting and luminous IR galaxies,][]{Bastian2008}.  We show the results of these computations in the top 
  panel of Figure  \ref{fig:mstar}.  Clearly reducing the 
  number of initially massive GCs significantly suppresses the BBH merger rate; even the  $M_{\rm GC}^{\bigstar} = 10^6 M_{\odot}$ truncation reduces 
  the local merger rate to $6~\rm{Gpc}^{-3}\rm{yr}^{-1}$ while the $2\times 
  10^5M_{\odot}$ truncation reduces the local merger rate to $4~\rm{Gpc}^{-3}\rm{yr}^{-1}$. 
  Of course, it is believed that many of these most massive GCs formed many Gyrs 
  ago, only to spiral into the center of their respective galaxies due to 
  dynamical friction \citep[e.g.,][]{Gnedin1997}.  While this would suppress any in-cluster 
  mergers from these clusters at late cosmic times due to tidal disruption, many of the ejected BBHs would remain outside of the disrupting clusters, allowing them to merge many Gyr after their birth clusters have been 
  destroyed.  Since the mergers of ejected BBHs are the larger contributor to the 
  rates presented here, it is entirely likely that many of the BBHs from these massive clusters will 
  still contribute to the overall merger rate.

  To better quantify the contributions from massive GCs to the predicted merger 
  rate, we recompute our standard model using the same $M_{\rm GC}^{-2}$ CIMF, 
  but with progressively decreasing upper limits for the maximum GC mass.  This is shown at 
  the bottom of Figure \ref{fig:mstar}.  As the upper limit is decreased from 
  $10^7M_{\odot}$ to $10^6M_{\odot}$, the rate in the local universe decreases 
  roughly linearly from  $14~\rm{Gpc}^{-3}\rm{yr}^{-1}$ to 
  $5~\rm{Gpc}^{-3}\rm{yr}^{-1}$.  Although our analytic model does likely 
  overestimate the number of mergers for the most massive
  clusters (see Appendix B), we note that these clusters do not
  dominate the BBH merger rate computed here, ensuring that our results are
  robust to within a factor of $\sim2$.

\subsection{Cluster Disruption}

We have so far assumed the fiducial model of \cite{El-Badry2018}, with an additional multiplicative factor of 2.6 taken from their 
Appendix D.  This additional factor was shown, when combined with the cluster 
disruption model of \cite{Choksi2018}, to reproduce the present-day relationship 
between galaxy halo mass and GC mass in the local universe \citep{Harris2014}.  
However, there is still a large amount of uncertainty regarding GC formation 
and disruption.  To attempt to bracket this uncertainty, we perform two 
additional calculations.  In the first, we assume the standard model of 
\cite{El-Badry2018}, but with no cluster disruption (i.e.~without the additional 
factor of 2.6).  This model is obviously unphysical, since it allows all 
low-mass clusters to survive to the present day, whereas in reality such 
clusters will be destroyed by two-body evaporation or tidal stripping.  Because 
this model significantly underpredicts the number of massive GCs in the local 
universe, we consider it a highly conservative (if unphysical) lower limit on GC disruption.

  \begin{figure}[bth]
\centering
\includegraphics[scale=0.85, trim=0.in 0.1in 0in 0.0in, clip=true]{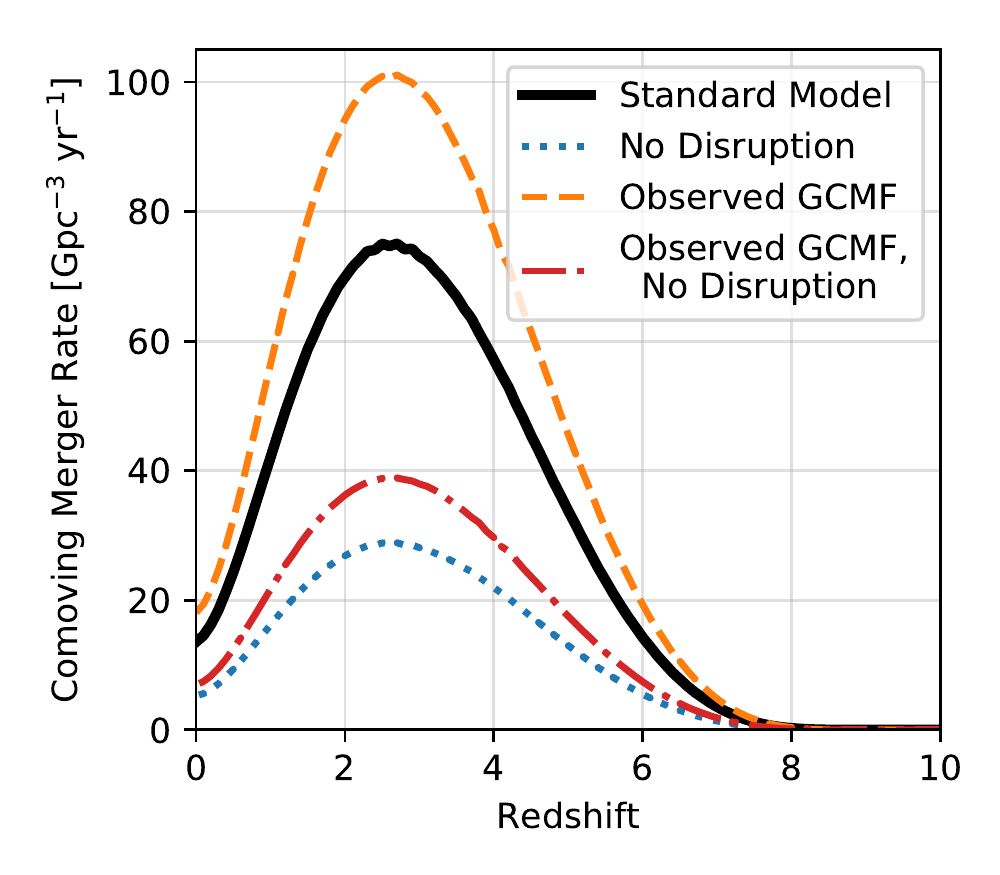}
\caption{Sensitivity of the BBH merger rate to assumptions about cluster disruption.  We show the merger rate computed assuming four different approaches to cluster disruption.  In the first two, we assume the standard $\propto M_{\rm GC}^{-2}$ CIMF with no disruption (highly nonphysical) in which all clusters survive for a Hubble time with no mass loss, and our standard model, which includes a correction factor from \citep[][]{El-Badry2018}.  We also show the merger rate assuming that GCs form according to the observed GCMF, with and without the correction factor for GC disruption.  }
\label{fig:disrupt}
\end{figure}

For the most optimistic assumption, we instead assume that clusters form according to the observed GC mass function (GCMF).  The observed mass distribution of present-day GCs has been shown to follow a 
roughly log-normal distribution \citep{Harris2014}, assuming a mass-to-light 
ratio of 2 \citep{Bell2003}, with a peak near $M_{\rm GC}^{\rm peak} = 3\times10^5 M_{\odot}$.  Dynamical modeling of GCs has 
shown that a typical GC near the peak of the present-day mass function 
 will lose approximately half its mass over 12 Gyr, 
largely due to stellar evolution, evaporation, and tidal stripping 
\cite[e.g.,][]{Morscher2015}.  To that end, we re-integrate Equation 
\eqref{eqn:master} using the log-normal luminosity function from 
\cite{Harris2014}, with the median increased by a factor of 4 (to account for 
both the mass-to-light ratio and the mass loss in an individual GC over $\sim 
12$ Gyr).  {Since this log-normal distribution is representative of GCs that 
survive disruption, we consider the GCMF with and without the additional factor 
of 2.6.  When the factor of 2.6 is included, this represented an upper bound on cluster disruption (since the observed GCMF has already been shaped by cluster disruption).  When it is not included, our model reverts to using the present-day observed population of GCs \citep[as was done in][]{Rodriguez2016a}, albeit with a different distribution of GC formation times.}

In Figure \ref{fig:disrupt}, we show 
our fiducal model alongside models with no disruption and disruption around the GCMF.  Our model with no disruption only achieves 
a merger rate of $5~\rm{Gpc}^{-3}\rm{yr}^{-1}$ in the local universe.  We 
reiterate that his model is unrealistic, as it does not include any mass loss or tidal disruption of 
clusters, but we include it for 
completeness.  Our model pinned to the observed GCMF increases the merger rate 
by $\sim 40\%$ to $18~\rm{Gpc}^{-3}\rm{yr}^{-1}$ when the additional factor of 
2.6 is employed, although we consider this equally unrealistic, as it 
overpredicts the number of massive clusters.  Combining this estimate of our most 
 optimistic and realistic assumptions about GC disruption with the smallest estimate from the 
previous section (where the CIMF was truncated at $M_{\rm 
GC}^{\bigstar}=2\times10^5M_{\odot}$), we can confidently bracket the merger rate 
in the local universe as lying between 5 and 18~$\rm{Gpc}^{-3}\rm{yr}^{-1}$, 
with a reasonable value of 14~$\rm{Gpc}^{-3}\rm{yr}^{-1}$.  
{Finally, the model using the observed GCMF without the disruption correction 
from \cite{El-Badry2018} yields a local merger rate of 
$7~\rm{Gpc}^{-3}\rm{yr}^{-1}$.  This assumes that only the present-day 
population of GCs contribute to the merger rate, similar to the 
calculation of \cite{Rodriguez2015a,Rodriguez2016a}, and yields a similar result 
(with a small increase arising from the inclusion of post-Newtonian effects and 
a distribution of cluster formation times). }

\section{Comparing Different Merger Rates}
\label{sec:rates}

  \begin{figure}[t!]
\centering
\includegraphics[scale=0.82, trim=0.in 0.in 0in 0.0in, clip=true]{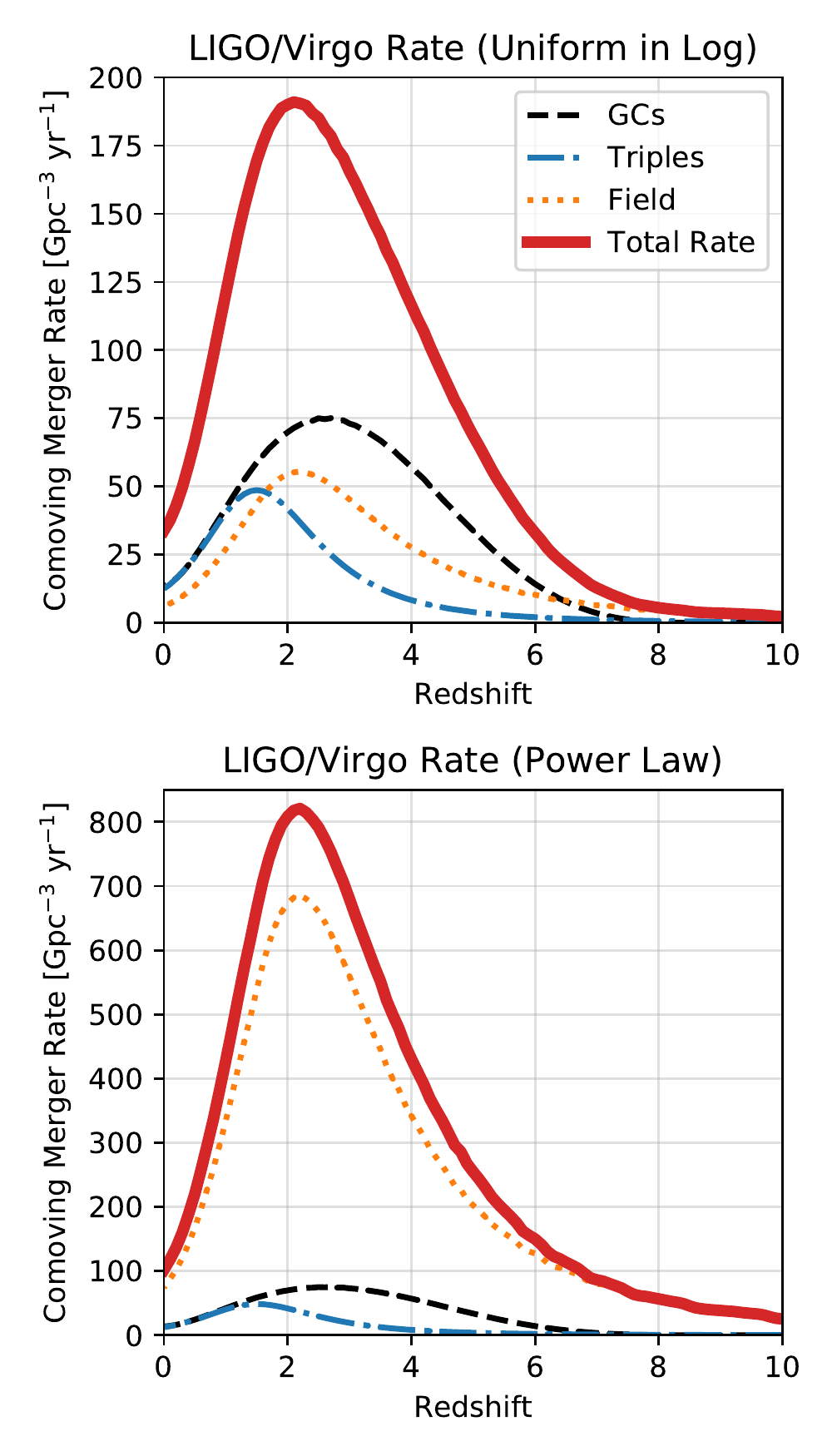}
\caption{Two examples of BBH merger scenarios, designed to reproduce the observed LIGO/Virgo merger rates (assuming a BH mass distribution either uniform in the logarithm or following a power law with a $-2.35$ index).  We show three different BBH formation channels: GCs, common-envelope evolution \citep[the field, taken from][]{Belczynski2016}, and mergers from field triples \citep[from][]{Rodriguez2018ALIGO/Virgo}.   In each case, the contribution from field binaries is adjusted to complete the observed LIGO/Virgo rate.}
\label{fig:rates}
\end{figure}

The fundamental question is whether any of these calculated merger 
rates can explain the BBH merger rate measured by Advanced LIGO/Virgo.  The 
current limits on the BBH merger rate in the local universe are model dependent 
and highly sensitive to the assumed BH mass distribution.  If a model with 
a uniform logarithmic mass distribution is assumed, then the current observed rates are 
$32^{+33}_{-22}~\rm{Gpc}^{-3}\rm{yr}^{-1}$ in the local universe at the 90\% 
level, fully 
consistent with the merger rates presented so far.  On the other hand, if a 
power-law BH mass distribution with a slope of $\alpha=-2.35$ is assumed (similar 
to the Kroupa slope for massive stars), then the rate increases to 
$103^{+110}_{-63}~\rm{Gpc}^{-3}\rm{yr}^{-1}$, which cannot be explained fully by the 
present analysis \citep[though we note that this estimate may be biased by the chosen upper-mass limit on BHs][]{Fishbach2017a}.

A full exploration of the many possible combinations of BBH formation channels 
with self-consistent physics is beyond the scope of this Letter.  
However, we can ask how three different BBH formation channels can combine to produce the observed LIGO/Virgo BBH merger rate.   For BBH mergers produced by the secular 
interactions with a third companion (the triples channel) we adopt the standard merger rates from \cite{Rodriguez2018ALIGO/Virgo}.  For BBH mergers via the classical common-envelope channel, we adopt the standard BBH merger rate from \cite{Belczynski2016}.  Rather than use multiple different realizations of the common-envelope channel, we adjust the overall normalization of the field BBH merger rate to whatever value would be required to fully explain the observed LIGO/Virgo merger rate (i.e. $\mathcal{R}_{\rm field} = \mathcal{R}_{\rm LIGO/Virgo} -  \mathcal{R}_{\rm GC} - \mathcal{R}_{\rm triples}  $ at $z < 0.1$)\footnote{Of course, we could have chosen any channel as the free parameter to yield the full LIGO/Virgo rate.  We have used the field channel for this purpose because it can potentially explain either all or none of the observed merger rate.}   We show these scenarios in Figure \ref{fig:rates}.

\begin{table}[bt]
\begin{tabular}{llll}
\toprule
\textbf{Model}                           & \textbf{$\mathcal{R}(z_{\rm max})/\mathcal{R}(0)$} & \textbf{$\mathcal{R}(1)/\mathcal{R}(0)$} & \textbf{$z_{\rm max}$} \\ \toprule
GCs (both $r_v$)                      & 5.5                                                 & 3.1                                       & 2.7                    \\
--- In-cluster & 8.8                                                & 4.1                                       & 2.9                    \\
--- Ejected    & 3.5                                                 & 2.4                                       & 2.3                    \\ \hline

GCs ($r_v = 1$pc)                      & 6.7                                                 & 3.5                                       & 2.7                    \\
--- In-cluster & 10.2                                                & 4.5                                       & 2.9                    \\
--- Ejected    & 4.3                                                 & 2.8                                       & 2.5                    \\ \hline
GCs ($r_v = 2$pc)                        & 4.0                                                 & 2.6                                       & 2.5                    \\
--- In-cluster & 6.8                                                 & 3.6                                       & 2.7                    \\
--- Ejected   & 2.5                                                 & 2.0                                        & 2.0                    \\ \hline
Triples                                  & 3.9                                                 & 3.2                                         & 1.5                    \\
Field                                    & 9.2                                                 & 4.5                                       & 2.2       \\  \toprule
\end{tabular}

\caption{The maximum merger rate (observable by $3^{\rm rd}$-generation detectors) and the rate at $z=1$ (observable by LIGO/Virgo) normalized by the merger rate in the local universe for the contribution from GCs, as well as the field and triple channels.}
\label{table}
\end{table}

 The standard dynamical assumptions produce a merger rate for GCs and field triples of 26 $\rm{Gpc}^{-3}\rm{yr}^{-1}$ in the local universe ($z < 0.1$).  If the log-uniform BH mass distribution is assumed to be the correct underlying distribution, then these three channels operate at approximately equal levels, with GCs contributing almost half of all mergers.  On the other hand, if the power-law mass function is assumed to be the correct distribution, then GCs contribute approximately 1 out of every 7 BBH mergers in the local universe (while triples contribute 1 out of every 9).  
 
 {We emphasize that the merger rates presented in Figure \ref{fig:rates} 
 represent only two possible scenarios, and that we have explicitly assumed that 
 the merger rate from GCs and from field triples are known.  In reality, we have 
 shown that the rate from GCs can easily span from 4 to 18 
 $\rm{Gpc}^{-3}\rm{yr}^{-1}$ in the local universe (while the rate from triples 
 is even less constrained).  This range may expand even further when a realistic distribution of GC initial radii is employed \cite[e.g.][]{Choksi2018a}.  For the log-uniform rate, this would imply that clusters may contribute anywhere from $\sim$ 1/8 to more than half of the observed BBH merger events, while the power-law rate would imply anywhere from $1/25^{\rm th}$ to $1/5^{\rm th}$ of BBHs are formed in GCs.   In reality, each of these channels contains significant systematic uncertainties, which are often correlated (e.g. the BH natal kicks, which can dramatically effect the rates from all three formation scenarios).}

\section{Discussion and Conclusion}

The most interesting aspect of the results presented in Section \ref{sec:rates} 
is that the peaks (and relative heights) 
of the merger rates from the different channels are unique.  This is hardly surprising, since the three physical mechanisms are 
expected to produce radically different delay time distributions.  In Table 
\ref{table}, we list the maximum of each merger 
rate and the rate at redshift $z=1$, normalized to the rate in the local universe.  Of all the merger rates analyzed here, 
the in-cluster merger rates peak earliest, at redshift 2.9, while triple-driven mergers 
peak latest at redshift 1.5.  Proposed 
third-generation detectors, such as LIGO Voyager or Cosmic Explorer \citep{Abbott2017f,Hild2011}
will be able to measure BBH mergers out to redshifts beyond 10.  At the same time, it has been shown that LIGO/Virgo may be able to measure the evolution of the BBH merger rate out to $z\sim 1$, and that this information may allow direct measurement of the BBH delay times within 2-5 years \citep{Fishbach2018}.  Although the growth in the GC and field rates are identical, the growth between the in-cluster or ejected mergers are significantly different, and may allow these channels to be distinguished by comparing detailed predictions for the masses and eccentricities of in-cluster and ejected BBH mergers from GCs.

Our results show a moderate enhancement over our \citep{Rodriguez2016a} previous estimates for the merger rate, which bracketed the rate between 2 and 20 $\rm{Gpc}^{-3}\rm{yr}^{-1}$, with a typical value of 5.  This increase arises from two factors: first, our newest models \citep{Rodriguez2018} include full post-Newtonian physics for BBH encounters inside the cluster, yielding a nearly 25\% increase in the merger rate, and a significant number of in-cluster mergers which were not present in previous studies \citep{Rodriguez2016a,Askar2016,Fragione2018,Hong2018}.  Secondly, our spatial density of GCs at high redshifts is somewhat larger than the assumed constant $\rho_{\rm GC} = 0.77 \rm{Mpc}^{-3}$ in \cite{Rodriguez2015a}.  This is because many GCs have disrupted before the present-day, which was not accounted for there.  While low-mass GCs do not contribute significantly, the massive clusters can play a significant roll, since any ejected BBHs can merge well after the destruction of their parent clusters.

{
Our results are mostly consistent with (if slightly higher than) similar studies by \cite{Fragione2018} and \cite{Hong2018}.  However, we note that the former models all GC formation as occurring instantaneously at $z=3$, while the later uses GC models that do not include any in-cluster mergers or post-Newtonian effects.  While this study was being finalized, we were informed of a similar work by \cite{Choksi2018a}, which coupled the semi-analytic model for BBH mergers from \cite{Antonini2016a} to a detailed model of GC formation and disruption.  Although their GC models incorporate less physics than those presented here, this semi-analytic treatment allows for a complete exploration of the parameter space for GC formation and its implication for the BBH merger rate.  This includes the initial virial radii of clusters, which we have not analyzed in significant detail due to computational constraints. When a similar distribution for GC effective radii is assumed, they find good agreement between our results and the ones presented here.}

\acknowledgments{
We thank Kareem El-Badry and Nick Choksi for useful discussions. CR acknowledges support from the Pappalardo Fellowship in Physics at MIT.  This work was supported in part by the Black Hole Initiative at Harvard University, which is funded by the JTF Foundation.}

\bibliographystyle{aasjournal}
\bibliography{Mendeley}{}

\appendix

\section{GC Formation Fits}
\label{app:elbadry}

\begin{figure}[h]
\centering
\includegraphics[scale=0.99, clip=true]{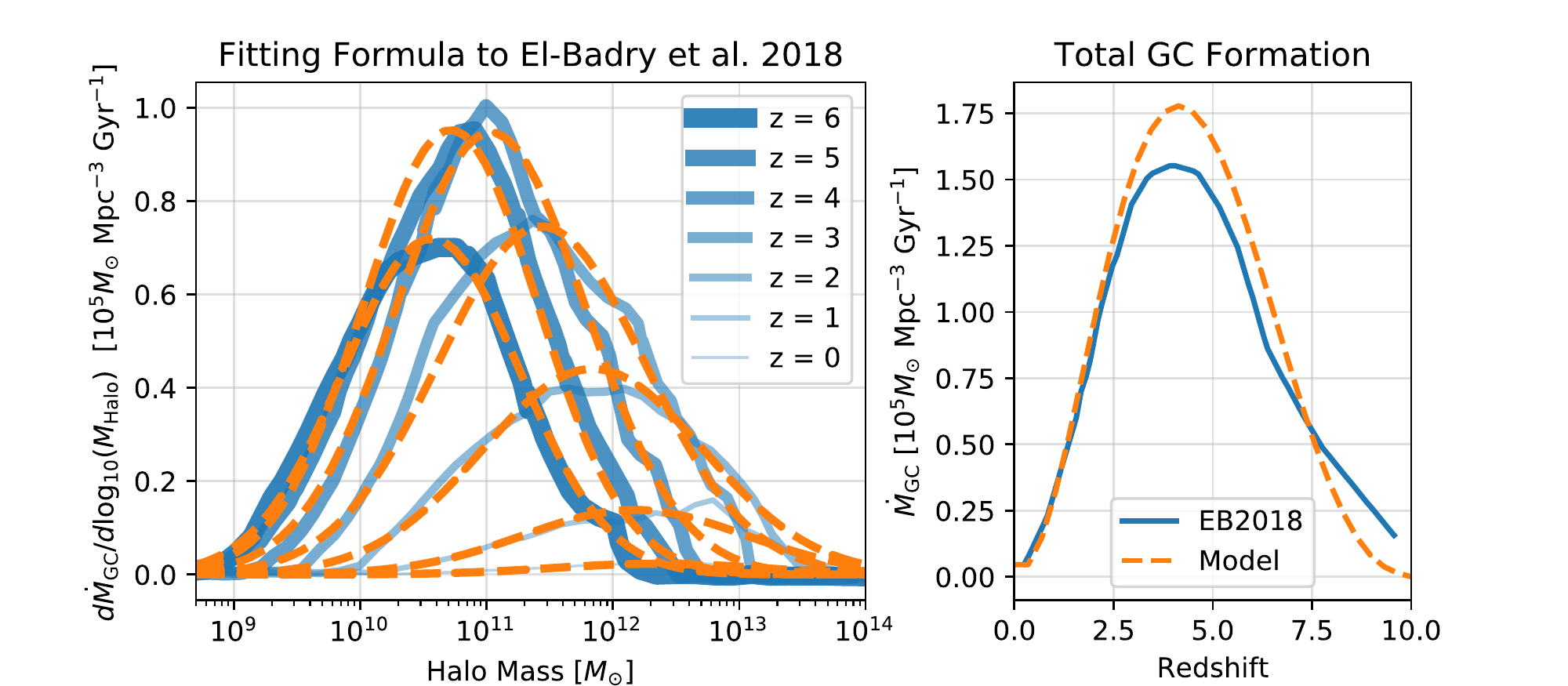}
\caption{The mass forming in GCs as a function of halo mass at different redshifts.  On the left, we show the original plot from \cite{El-Badry2018}, as well as our phenomenological fit in Equation \eqref{eqn:elbadryfits}.  The right shows the total GC formation rate as a function of redshift (found by integrating our fit over all halo masses), compared to the same quantity from \cite{El-Badry2018}.  We find that our predicted density of GCs in the local universe to be in good agreement when integrated over cosmic time ($5.8\times10^5M_{\odot}\rm{Mpc}^{-3}$ in our fit, versus $5\times10^5M_{\odot}\rm{Mpc}^{-3}$ in the original model).}
\label{fig:elbadry}
\end{figure}

For our fits to Figure 5 from \cite{El-Badry2018}, we require an analytic approximation to the cluster formation rate per year in different halo masses as a function of redshift.  This essentially forms the cosmological part of our computation, and takes the form of

\begin{equation}
\left. \frac{ \dot{M}_{\rm GC}} {d\log_{10}M_{\rm    
	 Halo}} \right|_{z(\tau)}
	\end{equation}
	
\noindent in units of $M_{\odot} \rm{yr}^{-1}\rm{Mpc}^{-3}$.  We find that a log-normal distribution of the form 

\begin{equation}
    \left. \frac{ \dot{M}_{\rm GC}} {d\log_{10}M_{\rm    
	 Halo}} \right|_{z} \approx \frac{A(z)}{\sqrt{2\pi}\sigma(z)} \exp\left( 
	 -\frac{\log_{10} M_{\rm halo} - \mu(z)}{2\sigma(z)^2} \right)
	 \label{eqn:elbadryfits}
\end{equation}

\noindent where $A(z)$, $\mu(z)$, and $\sigma(z)$ are fitted polynomials in the redshift $z$, fits their results well.  We show the original data from \cite{El-Badry2018} and our phenomenological fits in Figure \ref{fig:elbadry}.

As stated in the main text, we multiply Equation \eqref{eqn:elbadryfits} by an additional factor of 2.6, to account for cluster disruption.  The default model of \cite{El-Badry2018} does not include any mechanism for cluster disruption, and was designed primarily to reproduce the observed properties and halo mass/GC mass relationships observed in the local universe.  Of course, this meant that their model would unrealistically allow GCs with masses as low as $10^5M_{\odot}$ to survive to the present day.  To account for this, the authors applied the cluster disruption model of \cite{Choksi2018} to their model.  They found that to reproduce the observed GC mass/halo mass relationship in the local universe, they were required to increase their total GC formation rate by a factor of 2.6 \cite[see][Appendix D]{El-Badry2018}.

\section{Rate Fit}
\label{app:rate}

To generate our expression for $R(r_{v},M_{\rm GC},t)$, we use the GC models created in 
  \cite{Rodriguez2018}.  These models, created with the state-of-the-art Cluster 
  Monte Carlo code \citep{Joshi1999,Pattabiraman2013} contain all the necessary 
  physics to model the formation of merging compact objects, including detailed 
  stellar evolution \cite{Hurley2000,Hurley2002}, dynamical formation of 
  binaries from three isolated BHs \cite{Morscher2012}, dynamical encounters 
  between binaries and other single/binary stars \cite{Fregeau2008}.  Recently, 
  we have upgraded CMC to include fully post-Newtonian dynamics for BHs, 
  including GW emission for isolated binaries and during binary-single and 
  binary-binary encounters \citep{Rodriguez2018}.  This has greatly enhanced the
  number of BBH mergers which can occur in the cluster\footnote{These models are 
	  nearly identical to those presented in \cite{Rodriguez2018}.  However, 
	  an error in the relativistic physics during BH-BBH scatterings was discovered 
	  which reduced the number of in-cluster mergers presented in that work.  
	  We still find that $\sim 1/2$ of all BBH mergers occur inside the 
	  cluster, but this number reduces to $\sim 1/3$ in the local 
	  ($z < 1$) universe. This does not significantly change the results 
  quoted in that paper, but was sufficient to require the generation 
  of new models}, a significant deviation 
  over previous results \citep[e.g., ][]{Rodriguez2016a,Askar2016}.

  To fit the rate to each model, we count the number of mergers that 
  occur in 1 Gyr bins in each model of a given $M_{\rm GC}$.  Equation 
  \ref{eqn:gcrate} is then assumed to be the time-dependent rate for a Poisson 
  process, giving the probability of observing a certain number of mergers in a 
  bin of width $T$ at a given time $t$ from a cluster with initial mass mass $M_{\rm GC}$ as

  \begin{equation}
	  P(N | R(t, M_{\rm GC}, \theta),T) = e^{-R(t, M_{\rm GC}, \theta) T} 
	  \frac{(R(t, M_{\rm GC}, \theta)T)^N}{N!}
  \end{equation}
  
  \noindent where $\theta$ are the five adjustable parameters for Equation \ref{eqn:gcrate}.  

  Using the binned merger rate from our GC models, the likelihood for an observed merger rate given our model rate $R$ can be expressed as:

  \begin{equation}
\mathcal{L}(N | M_{\rm GC},t,\theta) \propto \prod_{i} P(N^i | R(t^i, M^i_{\rm GC}, \theta),T) 
  \end{equation}

  \noindent where $N^i$, $t^i$, and $M^i_{\rm GC}$ are the number of mergers in the bin of width $T$ at time $t$ from a GC of mass $M_{\rm GC}$.  The expression for the probability of $\theta$ is simply:
  
  \begin{equation}
    p(\theta | N,M_{\rm GC},t,T) \propto \mathcal{L}(N | M_{\rm GC},t,T,\theta) \times p(\theta)
    \label{eqn:posterior}
  \end{equation}
  
  \noindent where $p(\theta)$ is the prior probability on the parameters $\theta$.  We use a flat prior in the range:
  
  \begin{align*}
      A M_{\rm GC}^2 + B M_{\rm GC} + C &> 0 \\
     -\gamma - \gamma_M \log_{10}M_{\rm GC} &< 0
  \end{align*}
  
  \noindent for all $M_{\rm GC}$ between $10^5M_{\odot}$ and $10^7M_{\odot}$.  Using the \texttt{Emcee} package \citep{Foreman-Mackey2012}, we fit the four sets of merger rates (in-cluster versus ejected and 1pc verses 2pc) by minimizing the logarithm of Equation \eqref{eqn:posterior}.  This produces a set of four $\theta$ vectors for our rate fitting.  The overall merger rate for clusters with a given virial radius is simply the sum of the in-cluster and ejected rates.

  \begin{figure*}[h!]
\centering
\includegraphics[scale=0.85, trim=0.in 0.1in 0in 0.0in, clip=true]{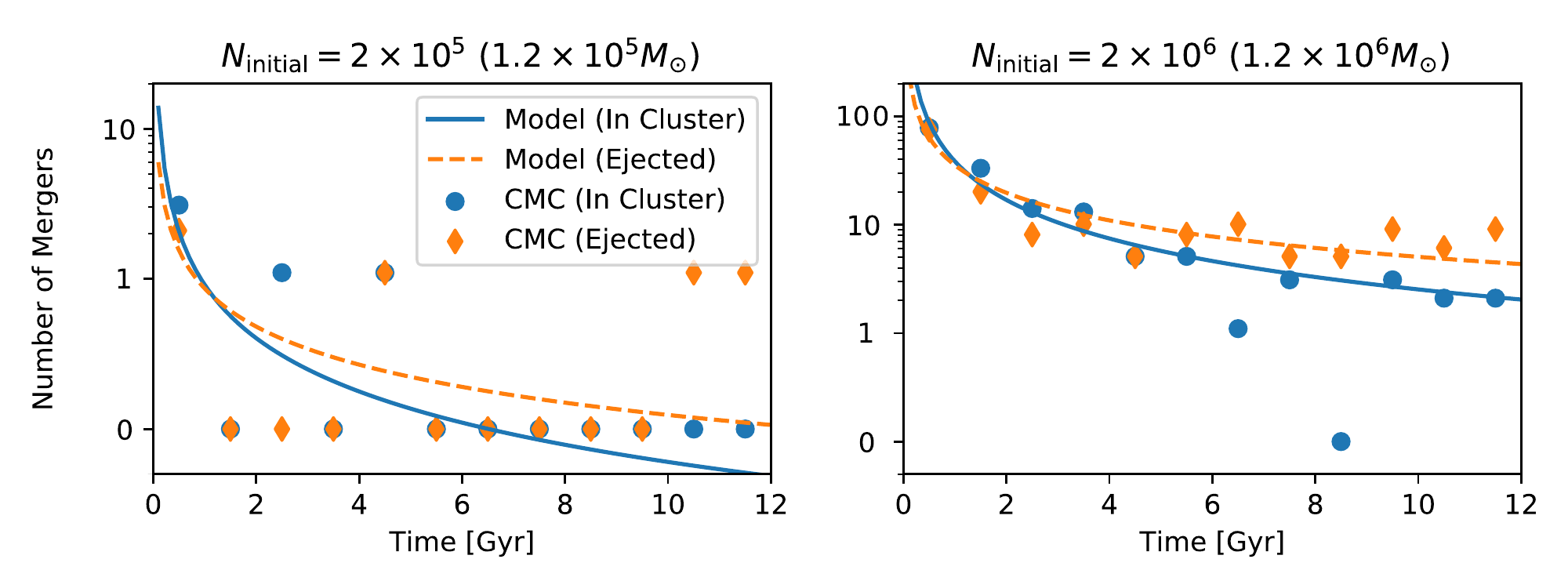}
\caption{The fit from Equation \eqref{eqn:gcrate} compared to the binned number of BBH mergers from two GC models \cite[from]{Rodriguez2018}.  We show two models with initial particle numbers of $2\times 10^5$ and $2\times10^6$.}
\label{fig:fits}
\end{figure*}
  
  As an example, we show the in-cluster and ejected mergers from two models with 
  $r_v = 1\rm{pc}$ in Figure \ref{fig:fits}.  We find that this function 
  reproduces well both the time-dependent merger rate and the variation with 
  cluster mass for all of our GC models.  However, we also note that our rate 
  function \eqref{eqn:gcrate} goes to infinity as $t\rightarrow 0$.  In reality, 
  the time required for mass segregation and the formation of BBHs means that 
  the first mergers often do not occur until $100$ Myr after cluster formation.  
  To account for this, we simply assume that \eqref{eqn:gcrate} goes to zero 
  when $t < 100$ Myr.  This increases the fidelity of our fit at early times.  
  We also found that the 100 Myr cutoff reduces the number of total mergers for 
  each GC (found by integrating \eqref{eqn:gcrate} over time from 0 to 12 Gyr) 
  to values that agree well with our CMC models.  We do note that this model may 
  over-predict the merger rate from the most massive GCs, since our fits predict 
  that a $10^7M_{\odot}$ GC may produce $\sim10^4$ mergers over 12 Gyr (in 
  contrast to other semi-analytic techniques, e.g. \cite{Antonini2016a}, where the number is closer to a few times $10^3$).  However, our CIMF largely disfavors the contribution from such large clusters.  We explore the implications of this in Section \ref{sec:3}.

\end{document}